\documentclass[conference]{IEEEtran}

\usepackage{epsfig,amsmath,amssymb,epsf,amsthm,scalefnt,multirow,subfig}
\usepackage{xcolor}
\usepackage{float}
\usepackage{cite}
\usepackage{psfrag}
\usepackage{mathtools}
\usepackage{graphics,amsmath,amssymb,graphicx,amsthm,scalefnt,multirow,dsfont,subfig}
\usepackage{latexsym,amsmath}
\usepackage{stfloats}
\usepackage{lipsum,cuted}	

\newcommand{\argmax}{\mathop{\mathrm{argmax}}}
\newcommand{\argmin}{\mathop{\mathrm{argmin}}}

\newtheorem{lemma}{Lemma}
\newtheorem*{lemma*}{Lemma}

\def\argmin{\mathop{\mathrm{argmin}}}
\def\argmax{\mathop{\mathrm{argmax}}}

\def\b0{{\pmb{0}}} 

\def\ba{{\mathbf{a}}}  \def\bc{{\mathbf{c}}} \def\bd{{\mathbf{d}}}
\def\bee{{\mathbf{e}}} \def\bff{{\mathbf{f}}} \def\bg{{\mathbf{g}}} \def\bh{{\mathbf{h}}}

\def\bq{{\mathbf{q}}}   
 \def\bv{{\mathbf{v}}}  
\def\by{{\mathbf{y}}} \def\bz{{\mathbf{z}}}  

\def\bA{{\mathbf{A}}} \def\bB{{\mathbf{B}}}  \def\bD{{\mathbf{D}}}
 \def\bF{{\mathbf{F}}}  
\def\bI{{\mathbf{I}}}  \def\bK{{\mathbf{K}}} 
   
 \def\bR{{\mathbf{R}}}  
  \def\bW{{\mathbf{W}}} \def\bX{{\mathbf{X}}}
\def\bY{{\mathbf{Y}}} \def\bZ{{\mathbf{Z}}}

\DeclareMathOperator{\E}{\mathbb{E}}

\begin{document}
	
	\title{Beam Design for Millimeter-Wave Backhaul with Dual-Polarized Uniform Planar Arrays}
	
	\author{$^1$Sucheol Kim, $^1$Junil Choi, and $^2$Jiho Song\\
			\IEEEauthorblockA{
				$^1$Department of Electrical Engineering, Korea Advanced Institute of Science and Technology\\
				$^2$Department of Electrical Engineering, University of Ulsan\\
				Email: \{loehcusmik, junil\}@kaist.ac.kr, jihosong@ulsan.ac.kr}
	}	
	\maketitle
	
	\begin{abstract}
		This paper proposes a beamforming design for millimeter-wave (mmWave) backhaul systems with dual-polarization antennas in uniform planar arrays (UPAs). 
		The proposed design method optimizes a beamformer to mimic an ideal beam pattern, which has flat gain across its coverage, under the dominance of the line-of-sight (LOS) component in mmWave systems. The dual-polarization antenna structure is considered as constraints of the optimization. Simulation results verify that the resulting beamformer has uniform beam pattern and high minimum gain in the covering region.
	\end{abstract}
	
	
	\section{Introduction}\label{sec1}

	Cell densification is a promising way to support the exponentially growing mobile devices and data rates 
	\cite{V.Changdrasekhar:2008,J.G.Andrews:2012,A.Ghosh:2012}
	. As cells become dense, the number of backhaul links increase, which would cause more frequent handover \cite{R.Taori:2015,J.G.Andrews:2013}. The resulting high data rate requirements for backhauls can be supported by conventional optical fibers, but it would be highly expensive to construct lots of backhaul links with optical fibers. A simple and cost effective backhaul solution is to use the millimeter-wave (mmWave) wireless communications that can support high data rates\cite{S.Hur:2013,X.Ge:2014}.
	
	Millimeter-wave (mmWave) communications assure enormous data rates with its huge bandwidth \cite{T.S.Rapport:2011}, if the high attenuation problem in the mmWave band could be resolved. An effective way of mitigating the attenuation is adopting sharp beam patterns, which can concentrate signal power. To reap the full benefit of the mmWave communications, hence, beamforming is essential. 
	Due to the dominant line-of-sight (LOS) component of mmWave channel \cite{J.Song:2017,A.Alkhateeb:2014}, the beamforming design problem can be considered as a graphical or geometrical shaping. The beamformings in \cite{S.Hur:2013} and \cite{L.Chen:2011}, for example, find the best beamformer by gradually shrinking the beamwidth of potential beamformers using predefined hierarchical codebooks. 

	The small wavelength of mmWave allows the beamforming with large number of antennas even under a small form factor. The cost and power of RF chains, however, cause the digital beamforming infeasible for mmWave \cite{O.E.Ayach:2014,C.H.Doan:2004}.
	A feasible solution is the analog beamforming, but its constant modulus condition reduces the diversity of beam pattern shape.
	As a compromise, the hybrid beamforming, which combines the digital and analog beamformings, are frequently adopted to balance both the feasibility and variety of beam pattern shape \cite{A.Alkhateeb:2014,A.Alkhateeb:2015}.
	
	By using dual-polarization antennas, additional increase of the number of antennas is possible within the same form factor. As a cost of doubled antennas, the beamforming for dual-polarization antennas should consider additional characteristics of dual-polarization channels \cite{B.Clerckx:2008}. 
	Most of previous dual-polarization beamformings, however, are only based on the digital beamforming.
	
	In this paper, we propose a hybrid beamforming design for mmWave backhaul systems with the dual-polarization antennas in uniform planar arrays (UPAs), which, to the best of authors' knowledge, has not been considered before. The ordinary backhaul links use predefined beamformers; however, as cells become dense, the number of new installations, demolitions, or movements of base stations (BSs) would increase, where each event necessitates new beamformers.
	A simple beamforming method would rely on discrete Fourier transform (DFT) codebook, but it is not straightforward to use the DFT codebook for the dual-polarization antennas in UPAs. Therefore, we first define an ideal beam pattern to have flat beamforming gain across its beam coverage, ensuring quality of service (QoS) with its high minimum gain. Then, we design a beamformer to have the most similar beam pattern with the ideal beam pattern. The similarity of two beam patterns are assessed by squared error (SE), and the dual-polarization UPA and hybrid beamforming structures are considered as constraints. Numerical results show that the proposed algorithm can generate uniform beam patterns with higher minimum gain than the previous beamforming method in \cite{B.Clerckx:2008}.
	
	$\textbf{Notations}$: Matrices and vectors are written in boldface capital and small letters $\bA$ and $\ba$. $(\cdot)^\mathrm{T}$, $(\cdot)^\mathrm{H}$, and $(\cdot)^*$ mean transpose, Hermitian, and element-wise conjugate of the corresponding matrix or vector. The Kronecker product and the Hadamard product are represented as $\otimes$ and $\odot$, respectively. 
	$\bI_a$ is the $a\times a$ identity matrix, $\bee_{a,b}$ is the $b$-th column of the identity matrix $\bI_a$, and $\boldsymbol{1}_a$ represents the $a\times1$ all one vector. The concatenation of matrices is denoted as $[\bA, \bB]$ where $\bA$ and $\bB$ have the same number of rows.

	\begin{figure}[t]
		\centering
		\subfloat[UPA deployment]{
			\includegraphics[width=.45\linewidth]{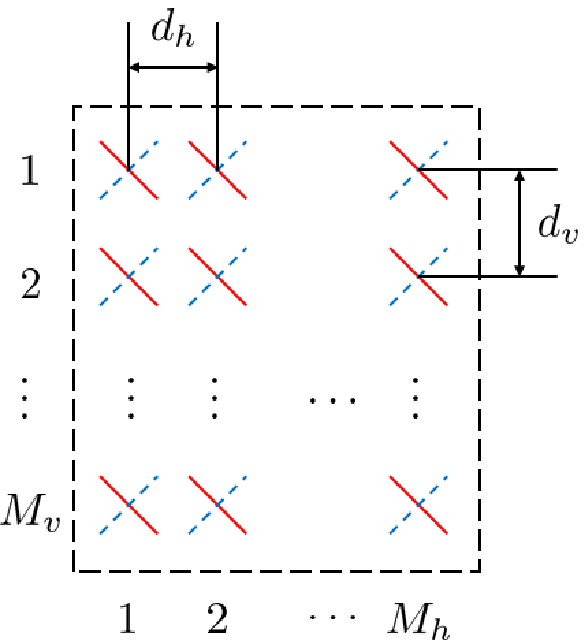}
			\label{array}
		}
		\subfloat[orientation difference $\phi$]{
			\centering
			\includegraphics[width=.4\linewidth]{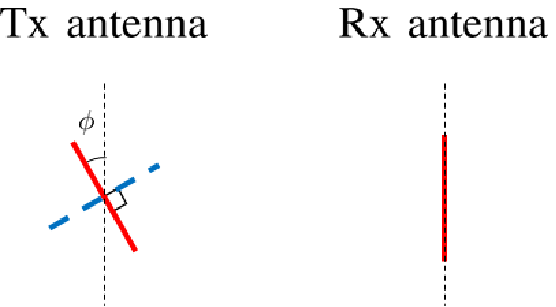}
			\label{orientation}
		}
		\caption{Dual-polarization antennas deployed in UPA.}
	\end{figure}
	
	\section{System and channel models}\label{sec2}
	\subsection{System model}\label{sec2-1}
	\begingroup
	\allowdisplaybreaks
	We consider multiple-input single-output (MISO) systems, where the extension to multiple-input multiple-output (MIMO) systems is possible with receive beamforming based on the proposed beam design approach. The transmitter is equipped with $M=2M_hM_v$ dual-polarization antennas in UPA as shown in Fig. \ref{array}. Antennas are fully connected to $N$ RF chains in the transmitter \cite{A.Alkhateeb:2014}.
	\endgroup
	With block fading assumption, a received signal can be modeled as
	\begin{align}\label{Rx signal}
	y=\sqrt{P} \bh^\mathrm{H} \bc s+n,
	\end{align}
	where $P\in\mathbb{R}$ is the transmit power, $\bh\in\mathbb{C}^{M\times1}$ is the channel vector, $\bc\in\mathbb{C}^{M\times1}$ is the unit-norm beamformer, $s\in\mathbb{C}$ is the data symbol with a constraint $\E[|s|^2]\le 1$, and $n\in\mathbb{C}$ is the additive white Gaussian noise (AWGN) with zero mean and variance $\sigma^2$. The signal-to-noise ratio (SNR) is ${P}/{\sigma^2}$. The beamformer $\bc$ is selected by the receiver within a codebook $\mathcal{C}=\{\bc^{(1,1)},\cdots,\bc^{(Q_h,Q_v)} \}$, which has $Q=Q_hQ_v$ codewords. In the rest of the paper, we will use the terms \textit{codeword} and \textit{beamformer} interchangeably.
	
	On the basis of the beam alignment as in \cite{S.Noh:2017,S.Hur:2013}, the receiver finds the codeword with the highest received power
	\begin{align}\label{align}
	(\check{p},\check{q})=\argmax_{ \substack{(p,q)\in\{1,\cdots,Q_h\}\times\{1,\cdots,Q_v\}} } \left\lvert \sqrt{P} \bh^\mathrm{H}\bc^{(p,q)} +n^{(p,q)} \right\rvert^2,
	\end{align}
	where $n^{(p,q)}$ is the AWGN at the $(p,q)$-th beam training with zero mean and variance $\sigma^2$. The receiver feeds back the selected index to the transmitter, and the transmitter sets the beamformer as
	\begin{align}
	\bc=\bc^{(\check{p},\check{q})}.
	\end{align}

	Each codeword is fully connected hybrid beamformer and consists of a digital and an analog beamformers as
	\begin{align}
	\bc=\bF\bv,
	\end{align}
	where $\bF=[\bff_1,\cdots,\bff_N ]  \in\mathbb{C}^{M\times N}$ is the analog beamformer, and $\bv\in\mathbb{C}^{N\times1}$ is the digital beamformer. Each element of the analog beamformer corresponds to a phase shifter and can be written as $e^{j\tau}$ with some $\tau\in[0,2\pi)$. 
	Note that practical phase shifters rely on quantized phases; however, it is known that having four or more bits for phase quantization gives the beamforming performance close to the full resolution \cite{O.E.Ayach:2014}. Therefore, we assume the full resolution for phase shifters in this paper.

	\subsection{Channel model}\label{sec2-2}
	The MISO channel with dual-polarization can be modeled as \cite{L.Jiang:2007}
	\begin{align}\label{channel model}
	\bh=\sqrt{\frac{M_hM_vK}{1+K}} \bh_\mathrm{LOS}+\sqrt{\frac{M_hM_v}{1+K}} \bh_{\mathrm{NLOS}},
	\end{align}
	where $K$ is the Rician $K$-factor, $\bh_\mathrm{LOS}\in\mathbb{C}^{M\times1}$ is the LOS component, and $\bh_\mathrm{NLOS}\in\mathbb{C}^{M\times 1}$ is the sum of several non-line-of-sight (NLOS) components.
	The LOS component of dual-polarization channel can be written as \cite{L.Jiang:2007}
	\begin{align}\label{LOS component}
	&\bh_\mathrm{LOS}\notag\\
	&
	=\bR(\phi) \left\{
	\left(	\begin{bmatrix}	\sqrt{\frac{1}{1+\chi}} \\ \sqrt{\frac{\chi}{1+\chi}} \end{bmatrix}
	\otimes \boldsymbol{1}_{M/2} \right) 
	\odot \left( \begin{bmatrix} \zeta^{vv}\\ \zeta^{hv} \end{bmatrix}
	\otimes \ba(\theta_{\mathrm{az}},\theta_{\mathrm{el}} ) \right) \right\},
	\end{align}
	where $\chi$ is the cross-polarization discrimination (XPD) value, which defines the distinction ability between different polarization antennas, $\zeta^{vv}\in\mathbb{C}$ is the complex channel gain from $v$ (vertically polarized) transmit antenna to $v$ receive antenna, $\zeta^{hv}\in\mathbb{C}$ is the complex channel gain from $h$ (horizontally polarized) transmit antenna to $v$ receive antenna,
	$\ba(\theta_{\mathrm{az}},\theta_{\mathrm{el}} )\in\mathbb{C}^{\frac{M}{2}\times1}$ is the single path array response vector of UPA with the LOS azimuth angle $\theta_{\mathrm{az}}$ and elevation angle $\theta_{\mathrm{el}}$, and $\bR(\phi)=\begin{bmatrix}\cos\phi &-\sin\phi\\ \sin\phi&\cos\phi\end{bmatrix}\otimes\bI_{M/2}$ is the Givens rotation matrix with the orientation difference $\phi$ between the transmit and the receive antennas \cite{L.Jiang:2007,B.Clerckx:2008}. Fig. \ref{orientation} shows the orientation difference $\phi$ between dual-polarization transmit antennas and vertically polarized receive antenna. In this and next sections, we assume fixed antenna arrays of a backhaul system with a fixed orientation difference $\phi$, where the transmitter suppose to know the difference.

	The array response vector of UPA is
	\begin{align}
	&\ba(\theta_{\mathrm{az}},\theta_{\mathrm{el}})=\ba_h (\theta_{\mathrm{az}},\theta_{\mathrm{el}})\otimes\ba_v (\theta_{\mathrm{el}}),
	\end{align}
	where $\ba_h (\theta_{\mathrm{az}},\theta_{\mathrm{el}} )\in\mathbb{C}^{M_h\times1}$ is the array response vector of horizontally arranged ULA, and $\ba_v (\theta_{\mathrm{el}} )\in\mathbb{C}^{M_v\times1}$ is the array response vector of vertically arranged ULA. Specifically, two array response vectors are written as
	\begin{align}
	\ba_h (\theta_{\mathrm{az}},\theta_{\mathrm{el}} )	
	&= 
	\frac{1}{\sqrt{M_h}}
	[1,e^{j \frac{2\pi d_h}{\lambda}  \sin \theta_{\mathrm{az}} \cos \theta_{\mathrm{el}}  },
	\notag\\
	&\qquad\qquad~~
	\cdots,e^{j \frac{2\pi d_h}{\lambda} (M_h-1) \sin \theta_{\mathrm{az}} \cos  \theta_{\mathrm{el}}  } ]^\mathrm{T}, \\
	\ba_v (\theta_{\mathrm{el}}) &= \frac{1}{\sqrt{M_v}}[1,e^{j \frac{2\pi d_v}{\lambda}  \sin \theta_{\mathrm{el}} },
	\notag\\
	&\qquad\qquad~~
	\cdots,e^{j \frac{2\pi d_v}{\lambda} (M_v-1) \sin \theta_{\mathrm{el}}  } ]^\mathrm{T},
	\end{align}
	where $d_h$ and $d_v$ is the interval of the horizontal and vertical ULA, and $\lambda$ is the wavelength of the carrier frequency. In this paper, we set $d_h=d_v=\frac{\lambda}{2}$ for simplicity.

	Considering practical cell sectorization, we focus on an angle range $(\theta_{\mathrm{az}},\theta_{\mathrm{el}})\in((-\frac{\pi}{2},\frac{\pi}{2}),(-\frac{\pi}{4},\frac{\pi}{4}))$. The corresponding horizontal and vertical spatial frequencies, i.e., $\psi_h=\frac{2\pi d_h}{\lambda}  \sin \theta_{\mathrm{az}} \cos\theta_{\mathrm{el}}$ and $\psi_v=\frac{2\pi d_v}{\lambda}  \sin\theta_{\mathrm{el}}$, are bounded as
	\begin{align}
	-\pi < \psi_h < \pi \label{spatial frequency bound},\quad 
	-\frac{\pi}{\sqrt{2}} < \psi_v < \frac{\pi}{\sqrt{2}}.
	\end{align}
	We can deal with the angle range $((-\frac{\pi}{2},\frac{\pi}{2}),(-\frac{\pi}{4},\frac{\pi}{4}))$ indirectly by considering its corresponding spatial frequency range $((-\pi,\pi),(-\frac{\pi}{\sqrt{2}},\frac{\pi}{\sqrt{2}}))$. For the sake of simplicity, we handle unpaired spatial frequencies 
	\begin{align}
	\psi_h=\frac{2\pi d_h}{\lambda}  \sin\theta_{\mathrm{az}},\quad \psi_v=\frac{2\pi d_v}{\lambda}  \sin\theta_{\mathrm{el}}
	\end{align}
	and array response vectors
	\begin{align}
	&\bd_h (\psi_h )= \frac{1}{\sqrt{M_h}}\left[1,e^{j\psi_h},\cdots,e^{j\psi_h (M_h-1) } \right]^\mathrm{T},	\\
	&\bd_v (\psi_v )= \frac{1}{\sqrt{M_v}}\left[1,e^{j\psi_v },\cdots,e^{j\psi_v (M_v-1) } \right]^\mathrm{T}
	\end{align}
	as in \cite{J.Song:2017}.

	Due to the large Rician $K$-factor of mmWave channels \cite{L.Jiang:2007}, in the following section, we design beams considering the dominant LOS component $\bh_\mathrm{LOS}$ in \eqref{LOS component} while the simulation results in Section \ref{sec4} are based on the channel model in \eqref{channel model}.
	
	\section{Beamforming design}\label{sec3}

	\subsection{Preliminary for beamforming design}\label{sec3-a}
	
	In this paper, we optimize a beamformer based on the squared error (SE) between the beam pattern of the beamformer and the ideal beam pattern, which will be defined shortly. We first derive the optimal digital beamformer and apply the orthogonal matched pursuit (OMP) algorithm to obtain the final hybrid beamformer as in \cite{O.E.Ayach:2014,Y.C.Pati:1993}. 
	
	To design the beamformer, we first quantize the spatial frequency range into $Q_h\times Q_v$ regions. Then, each quantized region is represented as
	\begin{align}
	B^{(p,q)}
	&
	=\bigg\{(\psi_h,\psi_v ): -\pi+\frac{2\pi(p-1)}{Q_h} \le \psi_h < -\pi+\frac{2\pi p}{Q_h} ,\notag\\
	&\qquad -\frac{\pi}{\sqrt{2}}+\frac{2\pi(q-1)}{\sqrt{2} Q_v} \le \psi_v < -\frac{\pi}{\sqrt{2}}+\frac{2\pi q}{\sqrt{2} Q_v} \bigg\},
	\end{align}
	where $p\in\{1,\cdots,Q_h \}$, and $q\in\{1,\cdots,Q_v \}$. To support the entire region with the minimum number of codewords, we cover each quantized region by one of $Q=Q_h Q_v$ codewords. The ideal beam pattern for each quantized region is defined to have a positive equal gain inside the region and zero gain outside the region.

	Similar to the procedures in \cite{J.Song:2017}, we derive the equal gain of the ideal beam pattern by considering the expected data rate conditioned on $\lVert \bh\rVert_2^2$
	\begin{align}\label{data rate}
	R_\mathrm{data}&=\underset{\psi_h,\psi_v}{\E}\left[\log_2\left(1+\frac{P}{\sigma^2}|\bh^\mathrm{H}\bc|^2\right)\bigg| \lVert\bh\rVert_2^2\right]\notag\\
	&
	=\underset{\psi_h,\psi_v}{\E}\left[ \log_2\left( 1+\frac{P}{\sigma^2}\lVert\bh\rVert_2^2 g_\mathrm{ref} (\psi_h,\psi_v,\bc)\right)\bigg|\lVert\bh\rVert_2^2 \right],
	\end{align}
	where $g_\mathrm{ref} (\psi_h,\psi_v,\bc)$ is the reference gain of a beamformer $\bc$, which is defined as
	\begin{align}\label{reference gain}
	&g_\mathrm{ref} (\psi_h,\psi_v,\bc)					\notag\\
	&
	=\left| \left[b\bR(\phi)\left\{ \begin{bmatrix} \rho_{pv}\\ \rho_{ph} \end{bmatrix}\otimes \big(\bd_h (\psi_h )\otimes\bd_v (\psi_v ) \big) \right\} \right]^\mathrm{H} \bc \right|^2,
	\end{align}
	where $b=\left( |\rho_{pv}|^2+|\rho_{ph}|^2\right)^{-\frac{1}{2}}$ is the normalization term, $\rho_{pv}=\sqrt{\frac{1}{1+\chi}} \zeta^{vv}$ and $\rho_{ph}=\sqrt\frac{\chi}{1+\chi}\zeta^{hv}$ are the complex gains related to $v$ and $h$ transmit antennas.
	With the reference gain, we will derive two lemmas. Due to the space limitation, we omit the proofs of lemmas in this paper while the proofs can be found in \cite{S.Kim:2019}.
	\begin{lemma}\label{lemma 1}
		The integral of the reference gain $g_\mathrm{ref} (\psi_h,\psi_v,\bc)$ of any unit-norm beamformer $\bc\in\mathbb{C}^{M\times1}$ have bound as
		\begin{align} 
		\int_{-\pi}^\pi \int_{-\pi}^\pi g_\mathrm{ref} (\psi_h,\psi_v,\bc)d\psi_h  d\psi_v \le \frac{(2\pi)^2}{M_h M_v},
		\end{align} 
		where the equality holds when $\bc$ is the linear combination of the vectors $\left[\rho_{pv} \bee_{\frac{M}{2},\ell}^\mathrm{T},\rho_{ph} \bee_{\frac{M}{2},\ell}^\mathrm{T} \right]^\mathrm{T}$, $\ell\in\left\{1,\cdots,\frac{M}{2}\right\}$.
	\end{lemma}
	We use Lemma \ref{lemma 1} to derive the second lemma, which gives an upper bound of the date rate \eqref{data rate} and defines the ideal beam pattern achieving the upper bound. 
	
	\begin{lemma}\label{lemma 2}
		In the region $B^{(p,q)}$, the ideal beam pattern
		\begin{align}\label{ideal beam pattern}
		g^{(p,q)}_\mathrm{ideal} (\psi_h,\psi_v )=\begin{cases}
		\frac{Q\sqrt{2}}{M_h M_v },   &(\psi_h,\psi_v )\in B^{(p,q)}    \\0,      &(\psi_h,\psi_v )\notin B^{(p,q)} 	\end{cases}
		\end{align}
		achieves the upper bound of the expected data rate \eqref{data rate}
		\begin{align}\label{upper bound}
		R_\mathrm{data}^\mathrm{upper}=\log_2\left(1+\frac{P}{\sigma^2}  \lVert \bh\rVert_2^2  \frac{Q\sqrt{2}}{M_hM_v} \right) .
		\end{align}
	\end{lemma}
	Using the definition of the ideal beam pattern in Lemma 2, we design beamformers in the following subsection.

	\subsection{Beamformer design}\label{SE}

	To assess the SE of two beam patterns, we represent beam patterns in vector forms. By partitioning each quantized region into $L_h\times L_v$ lattice sections, the vector form takes the gain of each section as an element
	\begin{align}
	\label{beam pattern vector ideal}
	\bg^{(p,q)}_\mathrm{ideal}
	&=G \bee_{Q_h,p} \otimes \bee_{Q_v,q} \otimes \boldsymbol{1}_L,                \\
	\label{beam pattern vector codeword}
	\bg(\bc)
	&=\left\lvert \left\{b\bR(\phi)\left( \begin{bmatrix} \rho_{pv}\\ \rho_{ph} \end{bmatrix} \otimes \bD \right)\right\}^\mathrm{H} \bc \right\rvert^2   ,
	\end{align}
	where $\bg^{(p,q)}_\mathrm{ideal}$ is the ideal beam pattern vector of region $B^{(p,q)}$, $G=\frac{Q\sqrt{2}}{M_h M_v}$ is the equal gain of the ideal beam pattern, $p\in\{1,\cdots,Q_h \}$, $q\in\{1,\cdots,Q_v \}$, $L=L_hL_v$, $\bg(\bc)$ is the beam pattern vector of the codeword $\bc$, and $\bD=\bD_h\otimes\bD_v$ with
	\begin{align}
	&
	\bD_h=\bigg[\bd_h \left(-\pi+\frac{\pi}{Q_h L_h}\right),\bd_h \left(-\pi+\frac{\pi}{Q_h L_h}+\frac{2\pi}{Q_h L_h}\right),\notag\\
	&\qquad\quad\ \ \cdots,\bd_h \left(-\pi+\frac{\pi}{Q_h L_h} +\frac{2\pi (Q_h L_h-1)}{Q_h L_h}\right)\bigg], 
	\end{align}	\begin{align}
	\notag\\
	&\bD_v=\bigg[\bd_v \left(-\frac{\pi}{\sqrt{2}}+\frac{\pi}{\sqrt{2}Q_v L_v}\right),\notag\\
	&\qquad\quad\ \ 
	\bd_v \left(-\frac{\pi}{\sqrt{2}}+\frac{\pi}{\sqrt{2}Q_v L_v}+\frac{2\pi}{\sqrt{2}Q_v L_v} \right),\notag\\
	&\qquad\quad\ \ 
	\cdots,\bd_v \left(-\frac{\pi}{\sqrt{2}}+\frac{\pi}{\sqrt{2}Q_v L_v}+\frac{2\pi (Q_v L_v-1)}{\sqrt{2}Q_v L_v} \right)\bigg]   .
	\end{align} 
	The columns of $\bD$ are the concatenation of array response vectors, each of which directing one of $QL$ sections. By assessing the SE between the two vector forms, we find the optimal codeword as
	\begin{align}\label{dual opt}
	\bc^{(p,q)}_{\mathrm{dual}}=\argmin_{\bc\in\mathbb{C}^{M\times1}}\lVert\bg^{(p,q)}_\mathrm{ideal}-\bg(\bc)\rVert_2^2.
	\end{align}
	In this section, we focus on the region $B^{(1,1)}$ and use $\bc_{\mathrm{dual}}$ as a simple notation of the optimal beamformer $\bc^{(1,1)}_{\mathrm{dual}}$.

	Due to the absence of the closed form solution of \eqref{dual opt}, we rewrite the vector forms \eqref{beam pattern vector ideal} and \eqref{beam pattern vector codeword} in other forms as
	\begin{align}
	\bg^{(1,1)}_\mathrm{ideal}  	
	&=\left\{\sqrt{G}  \bee_{Q_h,1}\otimes \bq_{L_h}\otimes \bee_{Q_v,1}\otimes\bq_{L_v} \right\} \notag\\
	&\quad \odot\left\{\sqrt{G}  \bee_{Q_h,1}\otimes\bq_{L_h}\otimes\bee_{Q_v,1}\otimes\bq_{L_v} \right\}^*    \label{ideal vector},
	\\
	\bg(\bc)&=\left\{b\left( \begin{bmatrix} \rho_{pv}\\ \rho_{ph} \end{bmatrix} \otimes \bD \right)^\mathrm{H} \bR(\phi)^\mathrm{H} \bc\right\}\notag\\
	&\quad \odot\left\{b \left( \begin{bmatrix} \rho_{pv}\\ \rho_{ph} \end{bmatrix} \otimes\bD\right)^\mathrm{H} \bR(\phi)^\mathrm{H} \bc\right\}^*, \label{beamformer vector}
	\end{align}		
	where $\bq_{L_a}\in\mathbb{C}^{L_a\times1}$ is any vector satisfying $\bq_{L_a}\otimes\bq_{L_a}^*=\boldsymbol{1}_{L_a}$, $a\in\{h,v\}$. With the rewritten forms, we can make a substitution, which gives a suboptimal solution, for the objective function in \eqref{dual opt} as
	\begin{align} \label{SE opt}
	\bc_{\mathrm{dual}}
	&=\argmin_{\bc\in\mathbb{C}^{M\times1}} \Bigg\lVert \gamma \left\{b \left( \begin{bmatrix} \rho_{pv} \\ \rho_{ph} \end{bmatrix} \otimes \bD\right)^\mathrm{H} \bR(\phi)^\mathrm{H} \bc\right\} \notag\\
	&\quad -\left\{\sqrt{G} (\bee_{Q_h,1}\otimes\bq_{L_h}\otimes\bee_{Q_v,1}\otimes\bq_{L_v} ) \right\}\Bigg\rVert_2^2 ,
	\end{align} 
	where $\gamma\in\mathbb{C}$ is a normalization constant. The constant $\gamma$ is the number that leads the Wirtinger derivative \cite{R.Hunger:2007} of the objective function \eqref{SE opt} to be zero, i.e.,
	\begin{align} \label{gamma value}
	&
	\gamma
	=  \frac{ \left\{b \left( \begin{bmatrix} \rho_{pv} \\ \rho_{ph} \end{bmatrix} \otimes \bD\right)^\mathrm{H} \bR(\phi)^\mathrm{H} \bc\right\}^\mathrm{H}}
	{\left\lVert b \left( \begin{bmatrix} \rho_{pv} \\ \rho_{ph} \end{bmatrix} \otimes \bD\right)^\mathrm{H} \bR(\phi)^\mathrm{H} \bc\right\rVert_2^2}  \notag\\
	&
	\qquad\ \ \cdot \sqrt{G}(\bee_{Q_h,1} \otimes \bq_{L_h} \otimes \bee_{Q_v,1} \otimes \bq_{L_v} ).
	\end{align} 
	To simplify the problem, we can handle the effective codeword ${\bc'}=\bR(\phi)^\mathrm{H} \bc$ where $\bR(\phi)$ is the Givens rotation matrix in \eqref{LOS component}. Because $\bR(\phi)$ is a unitary matrix, the multiplication with the Givens rotation matrix recovers the original codeword $\bc={\left(\bR(\phi)^\mathrm{H}\right)} ^{-1} {\bc'}=\bR(\phi){\bc'}$, and the effective codeword ${\bc'}\in\mathbb{C}^{M\times1}$ also satisfies the unit-norm constraint. With $\gamma$ and the effective codeword, the objective function of \eqref{SE opt} is written as
	\begin{align}\label{SE max}
	&\argmax_{{\bc'}\in\mathbb{C}^{M\times1}} \Bigg|\sqrt{G} (\bee_{Q_h,1} \otimes \bq_{L_h} \otimes \bee_{Q_v,1} \otimes \bq_{L_v} )^\mathrm{H} \notag \\
	&\qquad \qquad \cdot\frac{\left\{ \left( \begin{bmatrix} \rho_{pv} \\ \rho_{ph} \end{bmatrix}\otimes \bD \right)^\mathrm{H}{\bc'} \right\} }
	{\left\lVert  \left( \begin{bmatrix} \rho_{pv} \\ \rho_{ph} \end{bmatrix}\otimes \bD \right)^\mathrm{H}{\bc'}  \right\rVert_2}\Bigg|^2.
	\end{align} 
	In the objective function \eqref{SE max}, we first consider the denominator
	\begin{align}\label{SE denominator}
	&\left\lVert  \left( \begin{bmatrix} \rho_{pv} \\ \rho_{ph} \end{bmatrix}\otimes \bD \right)^\mathrm{H}{\bc'}  \right\rVert_2^2 \notag\\
	&\quad 
	={\bc'}^\mathrm{H} \left\{\left( \begin{bmatrix} \rho_{pv} \\ \rho_{ph} \end{bmatrix}\otimes \bD \right)\left( \begin{bmatrix} \rho_{ph} \\ \rho_{pv} \end{bmatrix}\otimes \bD \right)^\mathrm{H} \right\} {\bc'}\notag\\
	&\quad
	={\bc'}^\mathrm{H} \bK{\bc'},
	\end{align} 
	where $\bK=\left( \begin{bmatrix} \rho_{pv} \\ \rho_{ph} \end{bmatrix}\otimes \bD \right)\left( \begin{bmatrix} \rho_{pv} \\ \rho_{ph} \end{bmatrix}\otimes \bD \right)^\mathrm{H}$. The last equation implies that the eigenvalues of $\bK$ is the major consideration. In terms of eigenvalues, we can consider two sets 
	$\Omega=\left\{ \left[-\rho_{ph}^* \tilde{\boldsymbol{\nu}}^\mathrm{T},\rho_{pv}^* \tilde{\boldsymbol{\nu}}^\mathrm{T} \right]^\mathrm{T}: \tilde{\boldsymbol{\nu}}\in\mathbb{C}^{\frac{M}{2}\times1} \right\}$ and $\Gamma=\left\{\left[ \rho_{pv}\tilde{\boldsymbol{\mu}}^\mathrm{T},\rho_{ph}\tilde{\boldsymbol{\mu}}^\mathrm{T} \right]^\mathrm{T}: \tilde{\boldsymbol{\mu}}\in\mathbb{C}^{\frac{M}{2}\times1} \right\}$. 
	The elements in $\Omega$ span half of the vector space of dimension $\mathbb{C}^{M\times1}$, and the structure $\left[-\rho_{pv},~ \rho_{ph}\right]^\mathrm{H}\otimes \tilde{\boldsymbol{\nu}}$ implies they are the eigenvectors of $\bK$ with zero eigenvalue. The elements in $\Gamma$ are always orthogonal to elements in $\Omega$ and span rest half of the vector space. Therefore, any codeword can be represented by the sum of two vectors, one from $\Gamma$ and the other from $\Omega$ as
	\begin{align} 
	{\bc'}=x\boldsymbol{\mu}+z\boldsymbol{\nu}, ~~ \boldsymbol{\mu}\in\Gamma,\ \boldsymbol{\nu}\in\Omega, 	  
	\end{align} 
	where $x\in\mathbb{C}$ and $z\in\mathbb{C}$ are to satisfy $\lVert x\boldsymbol{\mu}+z\boldsymbol{\nu} \rVert_2^2=1$. 
	
	With the representation of the effective codeword,
	the objective function \eqref{SE max} can be rewritten as
	\begin{align} \label{SE replace}
	&\max 
	\Bigg| 
	\sqrt{G} (\bee_{Q_h,1} \otimes \bq_{L_h} \otimes \bee_{Q_v,1} \otimes \bq_{L_v} )^\mathrm{H} \notag\\
	&\qquad~ \cdot \frac{\left( \begin{bmatrix} \rho_{pv} \\ \rho_{ph} \end{bmatrix} \otimes \bD \right)^\mathrm{H} (x\boldsymbol{\mu}+z\boldsymbol{\nu})  }{\left\lVert \left( \begin{bmatrix} \rho_{pv} \\ \rho_{ph} \end{bmatrix} \otimes \bD \right)^\mathrm{H} (x\boldsymbol{\mu}+z\boldsymbol{\nu}) \right\rVert_2}
	\Bigg|^2 	\notag\\
	&
	\max
	\Bigg|
	\sqrt{G} (\bee_{Q_h,1} \otimes \bq_{L_h} \otimes \bee_{Q_v,1} \otimes \bq_{L_v} )^\mathrm{H}\notag\\
	&\qquad\quad~ \cdot \frac{ \left( \begin{bmatrix} \rho_{pv}\\ \rho_{ph} \end{bmatrix} \otimes \bD\right)^\mathrm{H} \boldsymbol{\mu}} { 
		\left\lVert \left( \begin{bmatrix} \rho_{pv} \\\rho_{ph} \end{bmatrix}\otimes \bD \right)^\mathrm{H} \boldsymbol{\mu}\right\rVert_2 } 
	\Bigg|^2.
	\end{align}
	The last equality in \eqref{SE replace} shows that the optimal effective codeword becomes $\bc'=x\boldsymbol{\mu}$. Based on the structure of $\boldsymbol{\mu} = [\rho_{pv}\tilde{\boldsymbol{\mu}}^\mathrm{T}, \rho_{ph}\tilde{\boldsymbol{\mu}}^\mathrm{T} ]^\mathrm{T},\ \tilde{\boldsymbol{\mu}}\in\mathbb{C}^{\frac{M}{2}\times1}$, we can consider sufficient conditions $\lVert\tilde{\boldsymbol{\mu}}\rVert_2^2=1$ and $x=b$ (note that $b$ is the normalization term defined after \eqref{reference gain}) instead of the unit-norm constraint $\lVert{\bc'}\rVert_2^2=1$. With the sufficient conditions, the simplified objective function becomes
	\begin{align} \label{SE last}
		&\max_{\tilde{\boldsymbol{\mu}}\in\mathbb{C}^{\frac{M}{2}\times1}} \Bigg| \sqrt{G} (\bee_{Q_h,1} \otimes \bq_{L_h} \otimes \bee_{Q_v,1} \otimes \bq_{L_v} )^\mathrm{H}
			\notag \\
			&\qquad\qquad\quad 
		\cdot \frac{ \left( \begin{bmatrix} \rho_{pv}\\ \rho_{ph} \end{bmatrix} \otimes \bD \right)^\mathrm{H} [\rho_{pv}\tilde{\boldsymbol{\mu}}^\mathrm{T},\rho_{ph}\tilde{\boldsymbol{\mu}}^\mathrm{T} ]^\mathrm{T} }
		{\left\lVert \left( \begin{bmatrix} \rho_{pv}\\ \rho_{ph} \end{bmatrix} \otimes \bD \right)^\mathrm{H} [\rho_{pv}\tilde{\boldsymbol{\mu}}^\mathrm{T},\rho_{ph}\tilde{\boldsymbol{\mu}}^\mathrm{T} ]^\mathrm{T} \right\rVert_2 } \Bigg|^2 		\notag\\
		&
		\stackrel{(a)}{=}\max_{\tilde{\boldsymbol{\mu}}\in\mathbb{C}^{\frac{M}{2}\times1}} \Bigg| \sqrt{G} (\bee_{Q_h,1} \otimes \bq_{L_h} \otimes \bee_{Q_v,1} \otimes \bq_{L_v} )^\mathrm{H} 
			\notag\\
			&\qquad\qquad\qquad 
		\cdot \frac{(|\rho_{ph}|^2+|\rho_{pv}|^2)\bD^\mathrm{H}\tilde{\boldsymbol{\mu}}} {\lVert (|\rho_{ph}|^2+|\rho_{pv}|^2)\bD^\mathrm{H} \tilde{\boldsymbol{\mu}} \rVert_2} \Bigg|^2 \notag
		\\
		&
		\stackrel{(b)}{=}\max_{\tilde{\boldsymbol{\mu}}\in\mathbb{C}^{\frac{M}{2}\times1}} \left| \frac{\sqrt{G}(\bD_{h,1}\bq_{L_h}\otimes\bD_{h,1}\bq_{L_v})^\mathrm{H}\tilde{\boldsymbol{\mu}} }{\lVert\bD^\mathrm{H} \tilde{\boldsymbol{\mu}}\rVert_2 }\right|^2	,
	\end{align}
	where $\bD_{h,1}  = \bD_h\cdot (\bee_{Q_h,1}\otimes \bI_{L_h})$, and $\bD_{v,1}=\bD_v\cdot (\bee_{Q_v,1}\otimes\bI_{L_v})$. The equalities $(a)$ and $(b)$ are derived by the properties of the Kronecker product $(\bX\otimes\bY)^\mathrm{H}=\bX^\mathrm{H}\otimes\bY^\mathrm{H}$ and $(\bW\otimes \bX)(\bY\otimes \bZ)=(\bW\bY)\otimes(\bX\bZ)$. The last equation in \eqref{SE last} is the same with the reformulated objective function in \cite{J.Song:2017}, which is for the single-polarization beamformer. Hence, the same solution holds for both objective functions, and the optimal dual-polarization beamformer is obtained as
	\begingroup	\allowdisplaybreaks
	\begin{align} \label{SE final}
	\bc_{\mathrm{dual}}
	&=\bR(\phi) \bc_\mathrm{opt}'	\notag\\
	&=b\bR(\phi)[\rho_{pv}\tilde{\boldsymbol{\mu}}_\mathrm{opt}^\mathrm{T},\rho_{ph}\tilde{\boldsymbol{\mu}}_\mathrm{opt}^\mathrm{T} ]^\mathrm{T}  \notag \\
	&=b\bR(\phi)\left( \begin{bmatrix} \rho_{pv} \\ \rho_{ph}\end{bmatrix} \otimes\bc_{\mathrm{single}} \right),
	\end{align}
	\endgroup
	where $\bc_{\mathrm{single}}=\tilde{\boldsymbol{\mu}}_\mathrm{opt}=\frac{\bD_{h,1} \bq_{L_h}\otimes\bD_{v,1} \bq_{L_v}}{\lVert\bD_{h,1} \bq_{L_h}\otimes\bD_{v,1} \bq_{L_v} \rVert_2}$ is the optimal single-polarization beamformer in \cite{J.Song:2017}. 
	
	The optimal beamformer $\bc_{\mathrm{dual}}$ depends on $(\bq_{L_h},\bq_{L_v})$, and we reflect this dependency by calling the beamformer as the beamformer candidate with the notation $\bc_{\mathrm{dual}} (\bq_{L_h},\bq_{L_v} )$. Each candidate becomes a hybrid beamformer by applying the OMP-based algorithm \cite{O.E.Ayach:2014,Y.C.Pati:1993}, while we omit the details due to space limitation. 
	Among the hybrid beamformers corresponding to each pair $(\bq_{L_h},\bq_{L_v})$, we select the final hybrid beamformer that has the minimum SE.

	\begin{figure}[t]
		\centering
		\subfloat[proposed codeword (SE$=1.4291$)]{
			\includegraphics[width=.49\linewidth]{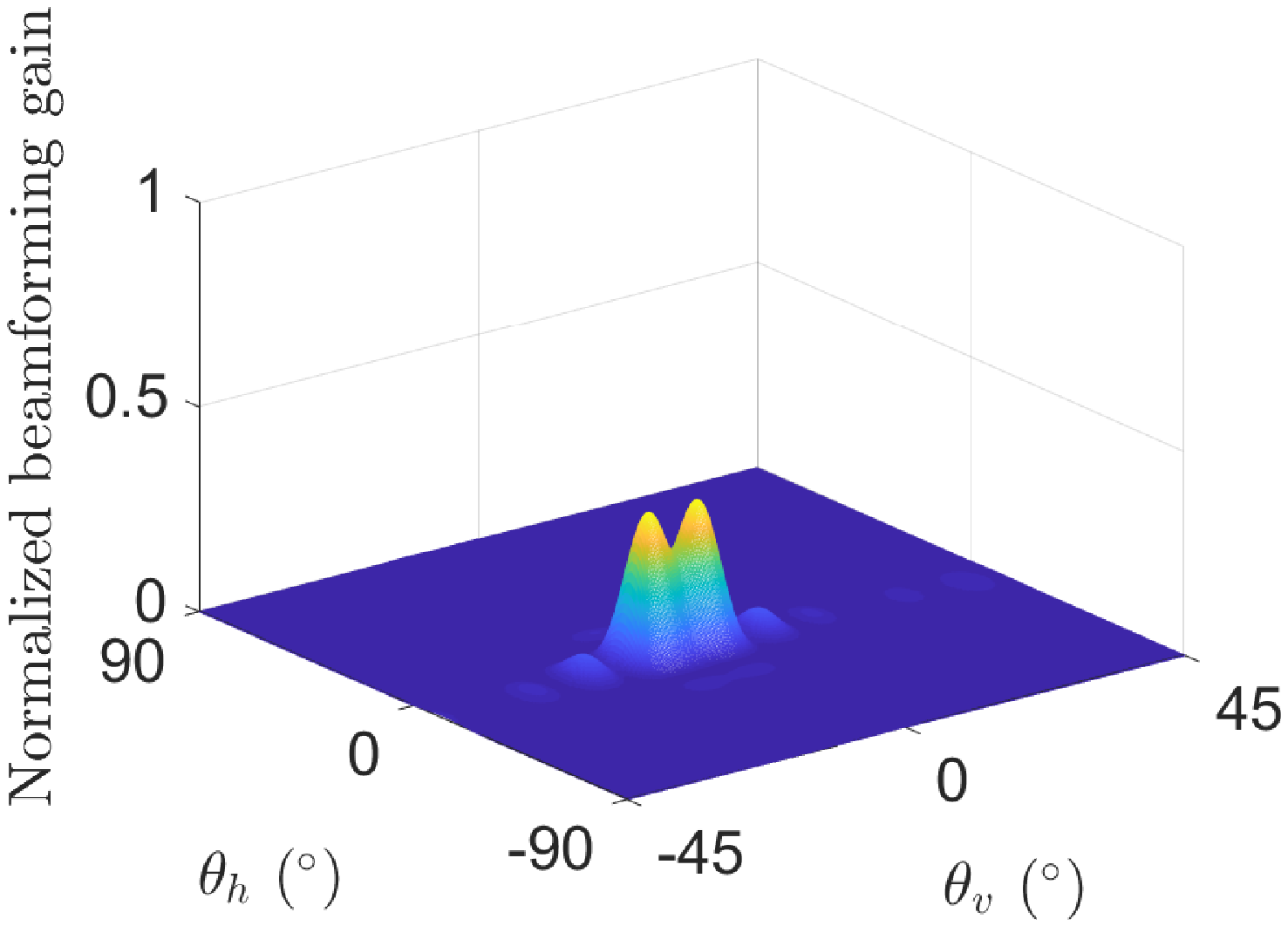}
			\label{SE codeword}
		}
		\subfloat[codeword in \cite{B.Clerckx:2008} (SE$=1.8668$)]{
			\includegraphics[width=.49\linewidth]{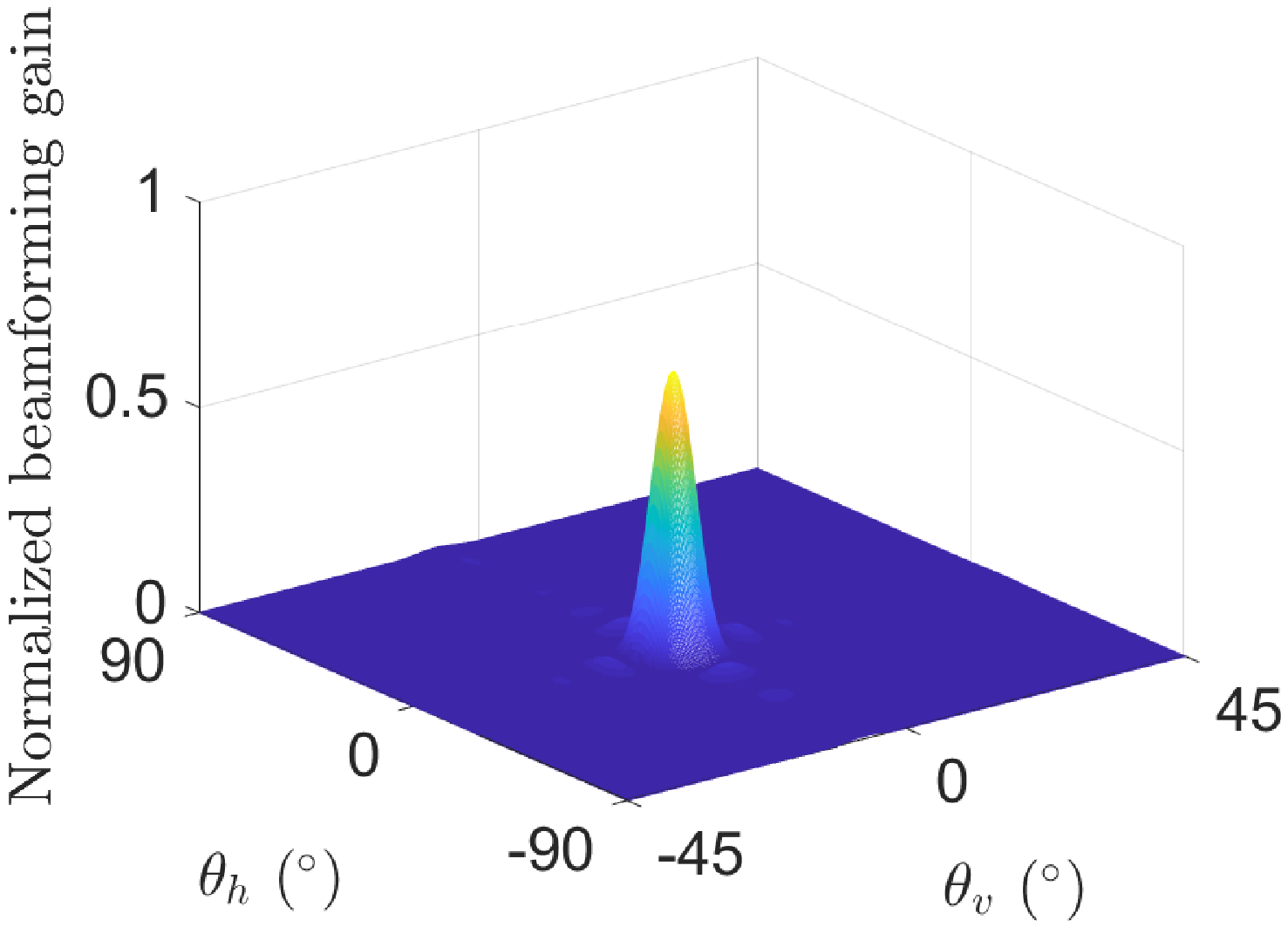}
			\label{RBD codeword}
		}
		\caption{Normalized beamforming gains at the region $B^{(3,3)}$ with $(M_h,M_v)=(8,16),\ (Q_h,Q_v)=(6,6)$.}
		\label{codeword}
	\end{figure}
	\begin{figure}[t]
		\centering
		\subfloat[{proposed codebook}]{
			\includegraphics[width=.49\linewidth]{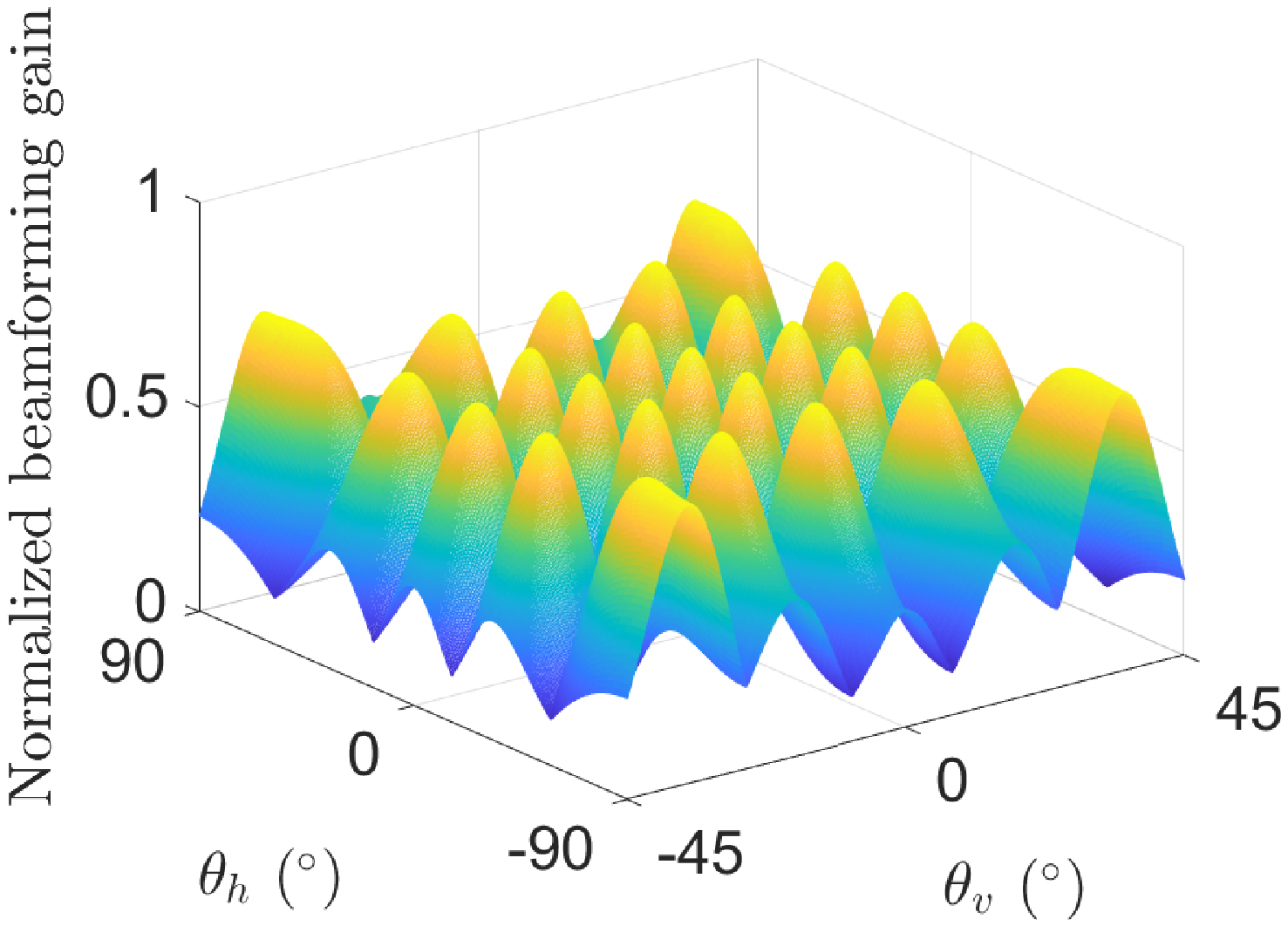}%
			\label{SE codebook}%
		}
		\subfloat[codebook in \cite{B.Clerckx:2008}]{
			\includegraphics[width=.49\linewidth]{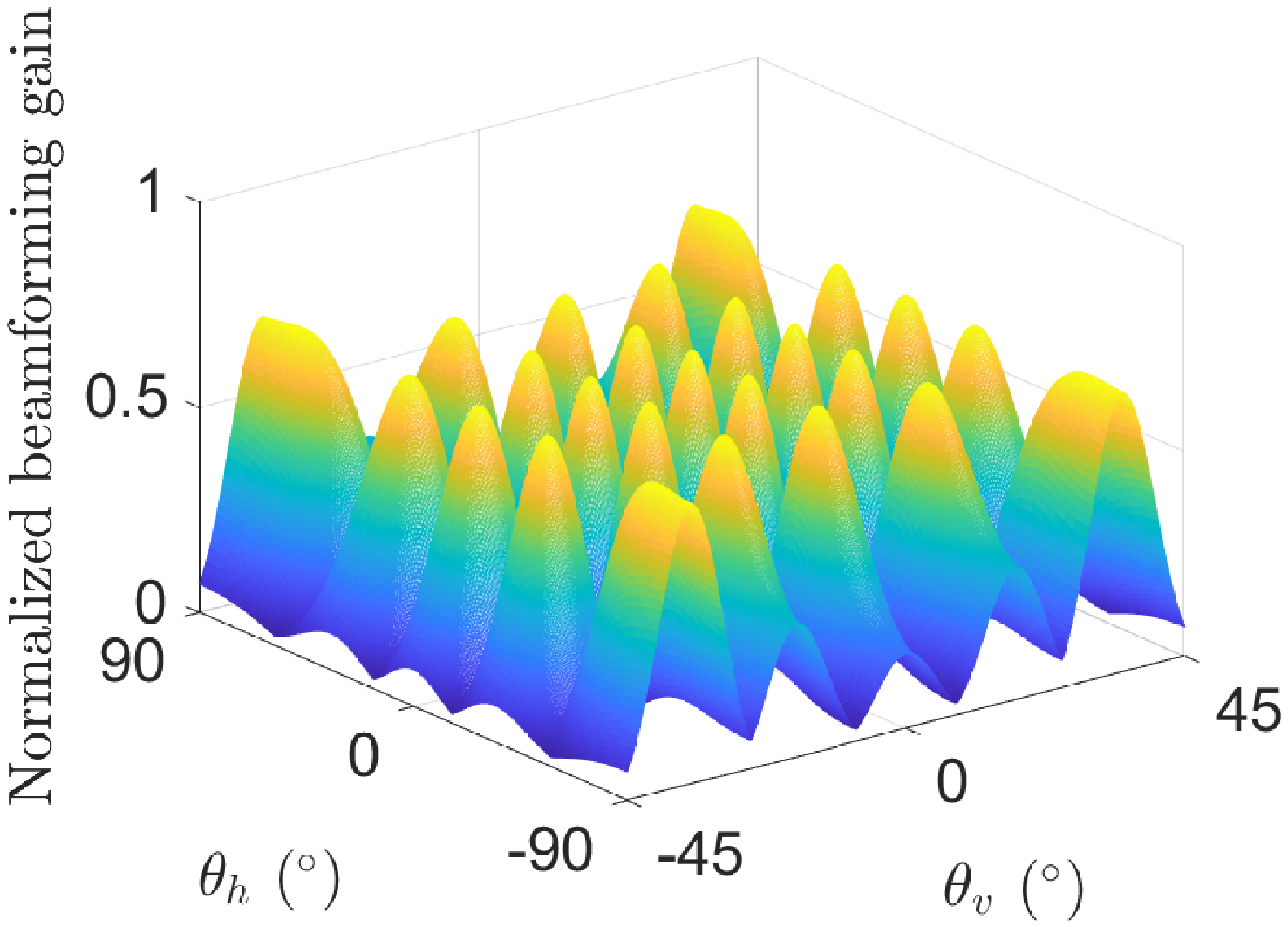}%
			\label{RBD codebook}%
		}
		\caption{Normalized beamforming gains with $(M_h,M_v)=(6,10),\ (Q_h,Q_v)=(5,5)$.}
		\label{codebook}
	\end{figure}
	
	\subsection{Channel information acquisition}\label{sec3-d}
	
	The proposed beamforming method requires the knowledge of the XPD value $\chi$, the orientation angle $\phi$, and the complex gains $\zeta^{vv}$ and $\zeta^{hv}$. With relative constancy of the XPD value, it is possible to assume $\chi$ is fixed and known to the transmitter. Since we focus on the backhaul systems, we assume $\phi$ is also known to the transmitter. 
	Regarding the channel gains $\zeta^{vv}$ and $\zeta^{hv}$, it is well known that the channel coherence time can be quite large after proper beam alignment in mmWave communications \cite{V.Va:2017}. Therefore, infrequent update of complex gains is sufficient, and we adopt a pilot-based method for the transmitter to obtain this information. For detailed steps, we refer to \cite{S.Kim:2019}.

	\section{Simulation results}\label{sec4}

	The numerical results of the proposed codebook are compared with that of the codebook in \cite{B.Clerckx:2008}. The transmitter is equipped with $N=4$-RF chains and $M=2M_h M_v$ dual-polarization antennas. In terms of the channel, we set the XPD value and the orientation difference as $\chi=0.3$ and $\phi=\frac{\pi}{4}$. The spatial frequency range is divided into $Q_h\times Q_v$ regions, each with $L_h\times L_v=7\times7$ sections. In the proposed codebook design, we consider the candidate set of $(\bq_{L_h},\bq_{L_v} )$ as
	\begin{align}
	\mathcal{G}_h\times \mathcal{G}_v=\left\{ (\by,\bz):y_i=e^{-\pi+\frac{2\pi}{B}\ell}, z_j=e^{-\pi+\frac{2\pi}{B} m}\right\},
	\end{align} 
	where $\ell\in\{1,\cdots,B\}$, $m\in\{1,\cdots,B\}$, $i\in\{1,\cdots,L_h \}$, and $j\in\{1,\cdots,L_v \}$ with $B=3$.

	\subsection{Beam pattern comparison}

	In Fig. \ref{codeword}, the codewords of the proposed codebook and the codebook in \cite{B.Clerckx:2008} are compared in terms of their beam patterns for the region $B^{(3,3)}$. The proposed codeword gives more uniform beam pattern with higher gain near the edge of the region than those of the codeword from \cite{B.Clerckx:2008}. This is due to the objective function of the proposed codebook, which uniformly distributes power over the region of interest. The codeword in \cite{B.Clerckx:2008} has higher peak gain, but the narrow beam pattern decrease the minimum gain in the covering quantized region. The same features of the two codebooks can be observed by their entire beam pattern in Fig. \ref{codebook}.

	\subsection{Data rate comparison}
	
	We consider the channel with the LOS component and three NLOS components. The Rician $K$-factor is $K=13.2$ dB, and the orientation difference, which can be affected from wind turbulence, is randomly chosen in $[\frac{\pi}{4} - \frac{\pi}{36}, \frac{\pi}{4} + \frac{\pi}{36}]$.
	The data rate is calculated as
	\begin{align} 
	R_\mathrm{rate}={\E}\left[\log_2 \left(1+\frac{P}{\sigma^2} |\bh^\mathrm{H} \bc|^2 \right) \right].
	\end{align} 
	
\begin{figure}[t]
		\centering
		\includegraphics[width=.845\linewidth]{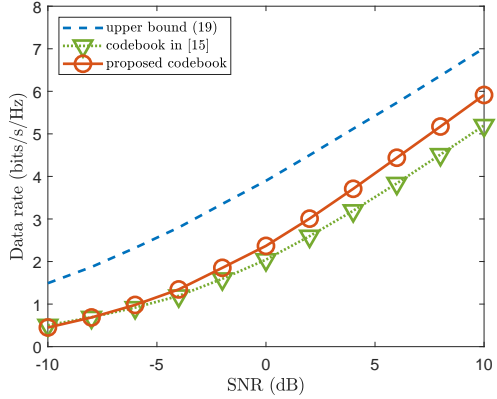}
		\caption{Data rate of codebooks with $(M_h,M_v)=(4,8),\ (Q_h,Q_v)=(5,4)$.}
		\label{data rate of double}
	\end{figure}

	In Fig. \ref{data rate of double}, the data rate of the proposed codebook is compared with that of the codebook in \cite{B.Clerckx:2008}, and the upper bound \eqref{upper bound} is presented as a reference. The data rate of the proposed codebook is higher than that of the codebook in \cite{B.Clerckx:2008} over the entire SNR. The gap between two data rates increases with SNR where the efficiency of beam pattern shape highly affects the performance.
	
	\section{Conclusion}\label{sec5}
	
	In this paper, we proposed the hybrid beamformer design method for mmWave backhaul systems equipped with dual-polarization antennas in UPA. The proposed beamforming design optimizes a codeword to generate a beam pattern similar to the ideal beam pattern, 
	while considering the dual-polarization UPA structure as optimization constraints. The proposed beam design outperforms the previous beam design based on the digital beamforming in \cite{B.Clerckx:2008}, which corroborates the efficiency of the proposed method. Although the proposed beamforming design is based on MISO system, the extension into MIMO system with multiple antennas at the receive BS is proposed in \cite{S.Kim:2019}.

	\section*{Acknowledgment}
	This work was supported by the National Research Foundation (NRF) grant funded by the MSIT of the Korea government (2019R1C1C1003638).
	
	\bibliographystyle{IEEEtran}
	\bibliography{references_bib}
	
\end{document}